\documentclass[prl,twocolumn,final,showpacs,showkeys,unsortedaddress]{revtex4}
\usepackage{graphics}

\begin{document}
\title{Amplification of evanescent waves in a lossy left-handed material slab}
\author{X. S. Rao} \email{tslrx@nus.edu.sg}
\affiliation{Temasek Laboratories, National University of Singapore, Singapore 119260}
\author{C. K. Ong}
\affiliation{Centre for Superconducting and Magnetic Materials and Department of Physics, National University of Singapore, Singapore 117542}
\date{\today}

\begin{abstract}
We carry out finite-difference time-domain (FDTD) simulations, with a specially-designed boundary condition, on {\it pure} evanescent waves interacting with a lossy left-handed material (LHM) slab. Our results provide the first full-wave numerical evidence for the amplification of evanescent waves inside a LHM slab of finite absorption. The amplification is due to the interactions between the evanescent waves and the coupled surface polaritons at the two surfaces of the LHM slab and the physical process can be described by a simple model. 
\end{abstract}

\pacs{42.25.Bs, 78.20.Ci, 42.30.-d, 73.20.Mf} 
\keywords{left-handed material (LHM), evanescent wave, surface polariton, superlens.} 

\maketitle

Recently, Pendry \cite{Pendry2000} suggested that a class of superlenses could be realized by a planar slab of the so-called left-handed materials (LHMs) possessing both negative permittivity and permeability. 
With such a superlens, not only the phases of propagating waves but also the amplitudes of evanescent waves from a near-field object could be restored at its image. Therefore, the spatial resolution of the superlens could overcome the diffraction limit of conventional imaging systems and reach subwavelength scale. Especially at the single frequency where $\epsilon (\omega)=\mu (\omega)=-1$, the superlens could even be {\it perfect} -- all the structural details of the object could be reconstructed at the image with unlimited resolution.  
It was soon realized that the superlens could be of great importance in different technological areas. 

While great research interests were initiated by the revolutionary concept, hot debates were also raised \cite{Hooft2001,Williams2001,Valanju2002,Garcia2002}. 
For example, based on the fact that LHMs are necessarily dispersive and dissipative \cite{Veselago1968,Landau1960},
Garc\'{i}a and Nieto-Vesperinas  \cite{Garcia2002}
argued that the unavoidable absorption in LHM, no matter how tiny it is, would suppress the amplification of evanescent waves and thus the 
{\it perfect} lens is only a theoretical artifact and does not exist in real world. 
Despite the great amount of analytic and numerical attempts \cite{Ziolkowski2001,Shen2002,Feise2002,Haldane2002,Fang2003,Gomez-Santos2003,Loschialpo2003,Cummer2003,Smith2003}, reported results are diverse and disputes still remain. 

It is the purpose of this paper to examine the feasibility of the superlensing effect of a realistic LHM slab using a full-wave finite-difference time-domain (FDTD) method. Since it is the evanescent waves responsible for the superlensing effect, our investigation concentrates on the interaction between evanescent waves and the LHM slab.  Our numerical results provide, for the first time, direct evidence that evanescent waves could be amplified inside a LHM slab of finite absorption. We  show how the amplification happens and what is the role of the surface polaritons, excited by evanescent waves at the two surfaces of the LHM slab. We also study the effects of absorption and the thickness of the LHM slab on the amplification. 

The major innovation introduced in this paper is the incorporation of a specially-designed boundary condition in the FDTD scheme to simulate {\it pure} evanescent waves interacting with a LHM slab. We are therefore able to explicitly study the characteristics of the evanescent waves and their dependence on different parameters.  
This is significantly different from previous works \cite{Ziolkowski2001,Loschialpo2003,Cummer2003}, where both propagating and evanescent waves were involved in the simulation and the behavior of evanescent waves could only be implied by the image obtained, which caused large discrepancies in the conclusions since many other factors might have also contributed to the image quality in the FDTD simulations. 

In this paper we consider a two-dimensional transverse electric (TE) case. A LHM slab in vaccum extends from $x=0$ to $L$ with surfaces normal to the $x$-axis and is infinite in the $yz$-plane. 
The LHM is modelled by causal permittivity ($\epsilon$) and permeability ($\mu$) with the identical plasmonic form,
\begin{equation}
\epsilon ( \omega ) = \mu ( \omega ) = 1 - \frac{\omega_p^2}{\omega^2 - i\omega \nu_c} ,
\end{equation}
where $\omega_p$ is the plasma frequency and $\nu_c$ is the collision frequency.  
The $z$-polarized electric field source is excited in the $x=-d$ ($d=L/2$ in our simulations) plane. The plane wave solution in vaccum is of the form, 
\begin{equation}
E_z (x,y,t) = E_{z0} e^{-i(k_x x+k_y y-\omega t)}, 
\end{equation}
where $k_x$ and $k_y$ are wave numbers along the $x$- and $y$-directions, respectively, and they satisfy the dispersion relation,
\begin{equation}
k_x^2+k_y^2=\frac{\omega^2}{c^2}.
\end{equation}
In the case of $k_y^2>k_0^2$ ($k_0^2=\omega^2/c^2$), $k_x$ is imaginary and corresponds to an evanescent wave along the $x$-direction.

Experimentally, pure evanescent waves can be generated by a prism using total internal reflection or a grating with a period finer than the wavelength of the radiation \cite{Kawata2001}. 
Numerically, this can be done more simply in the FDTD simulation with the help of a specially-designed boundary condition along the transverse direction.
Owing to the continuity of the field across the interfaces, the transverse wave number $k_y$ is a constant in the whole space, both inside and outside the LHM slab. Therefore, it is possible for us to explicitly assign a predefined $k_y$ for the simulated wave. One way to achieve this objective is to impose a boundary condition in $y$-direction. In this case, we require that the fields in the boundary satisfy $E_z(x,y\pm\Delta y)=E_z(x,y)e^{\mp ik_y\Delta y}$. This approach is an extension of the periodic boundary condition (PBC)\cite{Chan1995} where, normally, only the wave vectors within the first Brillouin zone are employed.

In the FDTD simulation, we apply the standard Yee's algorithm in which the leapfrog staggered electric field and magnetic field are updated in time using the two coupled Maxwell's curl equations \cite{Taflove2000}. The computational space is taken to be $4000\times 1$ cells with $\Delta x=\Delta y=0.3$ mm. 
At both ends in the $x$-direction, 10-cell uniaxial anisotropic perfectly-matched layer (UPML) \cite{Taflove2000} is added to truncate the computational space.
The time step used is $\Delta t=\Delta x /(2c)=0.5$ ps. A sinusoidal source with a smooth turn-on of 30 periods \cite{Ziolkowski2001} is employed to excite a monochromatic wave at the angular frequency $\omega\approx 11$ GHz. The corresponding wavelength in vacuum $\lambda_0$ (=$2\pi/k_0$) is about $566 \Delta x$. 
By choosing $\omega_p$ and $\nu_c$ in (1) appropriately, we have $\epsilon (\omega)=\mu (\omega)=-1-i\gamma$ ($\gamma$ denotes the loss term) in all our simulations.
The dispersive $\epsilon$ and $\mu$ is handled in time domain using the piecewise-linear recursive convolution (PLRC) method \cite{Taflove2000}.

In the simulation, we monitored the time evolution of $E_z$ at three different locations, the two surfaces of the LHM slab and the image plane. It was found that a transient process occurs at the begining of the simulation and after a certain period, the fields reach their steady state and vibrate sinusoidally with unchanged amplitudes. The time required to reach the final steady state and the dynamic characteristics of the time evolution vary dramatically with different  parameters ($\gamma$, $k_y$ and $L$). It is beyond the scope of this paper to describe the details of these simulations. Instead, we will only discuss the steady-state results obtained. 

\begin{figure}
\begin{center}
\includegraphics{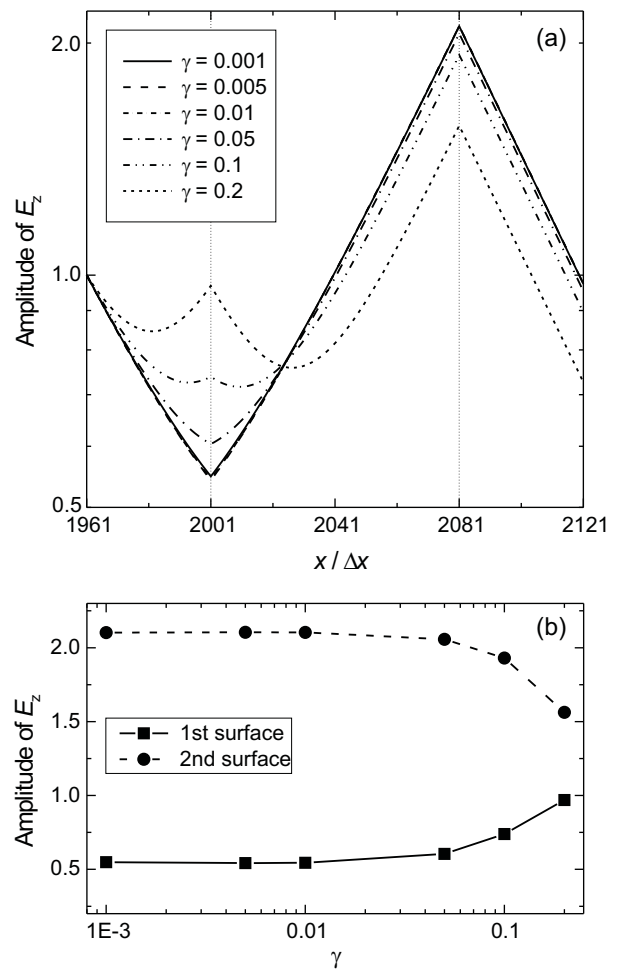}
\caption{An evanescent wave ($k_y=2k_0$) interacting with a LHM slab ($L=80\Delta x\approx 0.14\lambda_0$) for different values of absorption ($\gamma$), ranging from 0.001 to 0.2. (a) The $x$-distribution of the steady-state $E_z$ amplitude; (b) the steady-state $E_z$ amplitude at the two surfaces of the LHM slab as a function of $\gamma$. The vertical thin dot-lines in (a) denote the two surfaces of the LHM slab. }
\label{fig:1}      
\end{center}
\end{figure}

Figure 1(a) shows the spatial distribution of the steady-state amplitude of $E_z$ as a result of an evanescent wave ($k_y=2k_0$) interacting with a LHM slab of thickness $L=80\Delta x\sim 0.14\lambda_0$ for different values of $\gamma$, ranging from 0.001 to 0.2. Figure 1(b) shows the corresponding $E_z$ amplitude at the two surfaces of the slab as a function of $\gamma$.
The most important feature revealed by the figures is the amplification of the evanescent wave within the LHM slab, i.e., the field at the second surface is stronger than that at the first surface, for finite absorption, even though absorption tends to suppress the amplification. 
For small values of $\gamma$ from 0.001 to 0.01, the field distribution is almost identical. The evanescent wave from the source undergos very little reflection at the first surface of the LHM slab and the field between the source and the first surface is only slightly different from the exponential decay expected. Meanwhile, the field inside the slab shows a monotonic increase which is close to the exponential increase predicted by Pendry for the ideal case ($\epsilon=\mu=-1$) \cite{Pendry2000}.  
The field is amplified by about 4 times from 0.55 at the first surface to 2.1 at the second surface.    
When $\gamma$ increases above 0.01, the field at the first surface increases noticeably while that at the second surface drops, so that the amplification is weakened. When $\gamma$ reaches 0.2, the amplification still exists but now the field at the second surface (=1.56) is only about 1.5 times as strong as that of the first surface (=0.97). 
From the trends of the curves shown in Figure 1(b), we can expect that if $\gamma$ is further increased above a certain value $\gamma_c$($>0.2$), the field at the first surface may finally increase above the decreasing field at the second surface, so that the amplification would be replaced by decay.
Associated with the increase in the field at the first surface, larger reflections of the evanescent wave occur at the first surface. Furthermore, a peak appears at the first surface for $\gamma\geq 0.1$ and the peak becomes more prominent with increasing $\gamma$.  
 
\begin{figure}
\begin{center}
\includegraphics{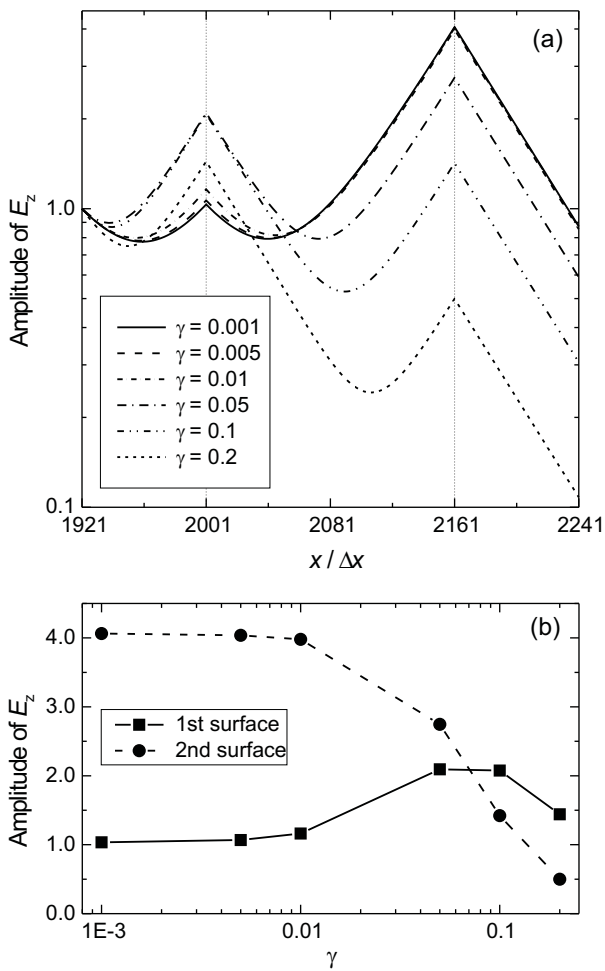}
\caption{An evanescent wave ($k_y=2k_0$) interacting with a LHM slab ($L=160\Delta x\approx 0.28\lambda_0$) for different values of absorption ($\gamma$), ranging from 0.001 to 0.2. (a) The $x$-distribution of the steady-state $E_z$ amplitude; (b) the steady-state $E_z$ amplitude at the two surfaces of the LHM slab as a function of $\gamma$. The vertical thin dot-lines in (a) denote the two surfaces of the LHM slab. }
\label{fig:2}      
\end{center}
\end{figure}

Figure 2 shows the simulation results for a thicker slab with $L=160\Delta x\sim 0.28\lambda_0$ while keeping $k_y=2k_0$ and $\gamma$ from 0.001 to 0.2. While the results show some common features with those obtained for the thinner slab presented in Figure 1, they also reveals a number of distinct features.
For $\gamma=0.001$, the LHM slab amplifies the field from 1.03 at the first surface to 4.06 at the second surface. The amplification is gradually weakened with increasing field at the first surface and decreasing field at the second surface while $\gamma$ is raised. At around $\gamma=0.07$ ($\gamma_c$), a crossover occurs where the fields at the two surfaces become equal.  
For $\gamma>0.07$, the field at the first surface is stronger than that at the second surface, and the amplification turns into decay. 
Compared with $\gamma_c>0.2$ obtained for the thinner slab, the crossover absorption value $\gamma_c=0.07$ is lowered dramatically for the thicker slab.  
The smaller $\gamma_c$ obtained for the thicker slab suggests that large $L$ is not favored for the amplification of the evanescent wave inside the LHM slab since it puts more stringent constraints.  
It is also clearly seen from Figure 2(a) that a peak is present, and large reflection occurs at the first surface, even for the lowest absorption studied ($\gamma=0.001$) where the LHM slab is supposed to be well matched to vacuum. This is different from the results shown in Figure 1(a). 

So far we have shown the simulation results of the evanescent wave interacting with LHM slabs of different absorption and thickness. The numerical examples provide direct evidence that an evanescent wave could be amplified in a LHM slab of finite absorption. 
As a result of the amplification of evanescent waves inside the lossy LHM slab, superlensing effect with subwavelength image resolution could be achieved practically \cite{Rao2003b}.  
It should be emphasized that our results are based on full-wave numerical experiments where only the casual $\epsilon(\omega)$ and $\mu(\omega)$ of the LHM are predefined and the results are obtained without any further assumptions and approximations.

Next, we will discuss the physical origin of the amplification of the evanescent waves inside the LHM slabs. It was shown by Ruppin that a vacuum/LHM interface supports surface polaritons (SPs), which are characterized by exponential decay in field on both sides of the interface \cite{Ruppin2000,Ruppin2001}. The peaks at the two surfaces of the LHM slabs and their shapes, shown in Figure 1(a) and 2(a), clearly illustrate the presence of SPs and emphasize their important roles in the systems. 
We believe that the coupled SPs, excited by the evanescent waves at the two surfaces of the LHM slab, are responsible for the amplification observed \cite{Haldane2002,Gomez-Santos2003,Pendry2002}. To prove this, we show that the features revealed by our numerical results are consistent with the predictions of a simple model of the interaction between the evanescent wave and the coupled SPs. 

The model we use to describe the interaction between the evanescent wave and the two coupled SPs is simply the same as the forced vibration and resonance of damped identical oscillators with linear coupling, which can be written in the coupled equations \cite{Gomez-Santos2003} 
\begin{eqnarray}
\ddot{\phi}_1 + \gamma\dot{\phi}_1+\omega_0^2 \phi_1 + \Omega_c^2 \phi_2 &=& F e^{i\omega t} \nonumber \\
\ddot{\phi}_2 + \gamma\dot{\phi}_2+\omega_0^2 \phi_2 + \Omega_c^2 \phi_1 &=& 0
\end{eqnarray}
where $\phi_1$ and $\phi_2$ denote the oscillations of the first SP and the second SP, respectively; $\omega_0$ is the natural resonant frequency of the single SP; $\Omega_c^2$ describes the coupling strength between the two SPs; and $F$ describes the external excitation introduced by the evanescent wave. 
When the evanescent wave is of frequency $\omega_0$, which is the case in our simulation, the steady-state solutions of $\phi_1$ and $\phi_2$ have the simple forms
\begin{eqnarray}
\phi_1 (\omega_0) =\frac{-i\gamma\omega_0 Fe^{i\omega_0 t}}{\Omega_c^4+ \gamma^2\omega_0^2}, \hspace{5mm}
\phi_2 (\omega_0)=\frac{\Omega_c^2 Fe^{i\omega_0 t}}{\Omega_c^4+ \gamma^2\omega_0^2}.
\end{eqnarray}
The characteristics of the coupled SPs and their dependences on different parameters can  be readily analyzed from these solutions.
In the lossless limit $\gamma\rightarrow 0$, the first SP disappears with $|\phi_1|=0$ and only the second SP is present in the system with finite coupling. This is the ideal case and an exponential amplification occurs inside the LHM slab. 
For finite absorption both SPs are excited. Simple calculation shows that, $|\phi_1|$ increases with increasing $\gamma$ until it passes through a peak, located at $\gamma=\gamma_p$ ($\gamma_p=\Omega_c^2/\omega_0$), and decreases beyond. On the other hand,  $|\phi_2|$ decreases monotonically with increasing $\gamma$. 
The ratio $|\phi_1/\phi_2|=\gamma\omega_0/\Omega_c^2$ gives a measure of the relative strength of the two SPs.
For small values of $\gamma$, $|\phi_1/\phi_2|<1$, the second SP is stronger than the first so that amplification takes place. However, the effect is reversed for $\gamma>\gamma_c$ ($\gamma_c=\Omega_c^2/\omega_0$) where the first SP is dominant. It is clear that $\gamma_c=\gamma_p$, which means that the crossover of $|\phi_1|$ and $|\phi_2|$ is accompanied by the peak in $|\phi_1|$. All these features are clearly observed in Figures 1 and 2. Furthermore, it is expected that increasing $L$ reduces $\Omega_c^2$, so that the value of $\gamma_c$ is smaller for the thicker LHM slab, which is also consistent with the numerical results.     

To explain why the first SP is stronger for the thicker slab when $\gamma$ is small and unchanged, we need to know the expressions of 
$\Omega_c^2$ and $F$ explicitly on the geometry. In the weak coupling limit $\kappa_x L\rightarrow \infty$ ($\kappa_x=\sqrt{k_y^2-\omega^2/c^2}$), they can be approximately written in the forms \cite{Haldane2002,Gomez-Santos2003}: $F\approx Ce^{-\kappa_x L/2}$ and $\Omega_c^2\approx D e^{-\kappa_x L}$, where $C$ and $D$ are coefficients independent of $\kappa_x L$.
Then, we have
\begin{eqnarray}
\frac{\partial |\phi_1(\omega_0)|}{\partial (\kappa L)} &=&\frac{1}{2}\frac{\gamma\omega_0 F}{(\Omega_c^4+ \gamma^2\omega_0^2)^2}(3\Omega_c^4-\omega_0^2\gamma^2). 
\end{eqnarray}
When $3\Omega_c^4-\omega_0^2\gamma^2>0$, i.e., $\gamma<\sqrt{3}\gamma_c$,
the first SP is getting stronger with increasing $L$, which explains the stronger first SP presented in Figure 2. It should be mentioned that the weak coupling limit may not be applicable for our simulations, where $\kappa_x L \approx 0.24$ (for $L=80\Delta x$) and 0.48 (for $L=160\Delta x$). However, it aids in understanding the result.
In fact, from a purely physical point of view, the emergence of the first SP with certain strength is crucial for the effective excitation of the second SP in the thicker slab, where the coupling coefficient $\Omega_c^2$ is reduced. It is the first SP that mediates the energy transfer from the evanescnet waves to the second SP.


In conclusion, we have proposed a method to simulate {\it pure} evanescent waves interacting with a LHM slab using FDTD. 
The simulation results provide the first full-wave numerical evidence that evanescent waves could be amplified inside a LHM slab with finite absorption. It also reveals that the coupled SPs, excited by the evanescent waves at the two surfaces of the LHM slab, play important roles in the amplification effect. We have shown that absorption tends to suppress the amplification and the amplification could be replaced by decay when the absorption is above the crossover value $\gamma_c$, which is greatly reduced with increasing thickness of the LHM slab. 
All features in our numerical results can be understood consistently using a simple model which describes the interaction between the evanescent waves and the coupled SPs.


\begin{thebibliography}{}
\footnotesize 
\bibitem{Pendry2000} J. B. Pendry, Phys. Rev. Lett. \textbf{85}, 3966 (2000).

\bibitem{Hooft2001} G. W. 't Hooft, Phys. Rev. Lett. \textbf{87}, 249701 (2001).

\bibitem{Williams2001} J. M. Williams, Phys. Rev. Lett. \textbf{87}, 249703 (2001).

\bibitem{Valanju2002} P. M. Valanju, R. M. Walser, and A. P. Valanju, Phys. Rev. Lett. \textbf{88}, 187401 (2002).

\bibitem{Garcia2002} N. Garcia and M. Nieto-Vesperinas, Phys. Rev. Lett. \textbf{88}, 207403 (2002).

\bibitem{Veselago1968} V. G. Veselago, Sov. Phys. Usp. \textbf{10}, 509 (1968).

\bibitem{Landau1960} L. D. Landau and E. M. Lifshitz, \textit{Electrodynamics of Continuous Media} (Pergamon Press, London, 1960).

\bibitem{Ziolkowski2001} R. W. Ziolkowski and E. Heyman, Phys. Rev. E \textbf{64}, 056625 (2001).

\bibitem{Shen2002} J. T. Shen and P. M. Platzman, Appl. Phys. Lett. \textbf{80}, 3286 (2002).

\bibitem{Feise2002} M. W. Feise, P. J. Bevelacqua, and J. B. Schneider, Phys. Rev. B \textbf{66}, 035113 (2001).

\bibitem{Haldane2002} F. D. M. Haldane, cond-mat/0206420.

\bibitem{Fang2003} N. Fang and X. Zhang, Appl. Phys. Lett. \textbf{82}, 161 (2003).

\bibitem{Gomez-Santos2003} G. G\'omez-Santos, Phys. Rev. Lett. \textbf{90}, 077401 (2003).

\bibitem{Loschialpo2003} P. F. Loschialpo, D. L. Smith, D. W. Forester, F. J. Rachford, and J. Schelleng, Phys. Rev. E \textbf{67}, 025602 (2003).

\bibitem{Cummer2003} S. A. Cummer, Appl. Phys. Lett. \textbf{82}, 1503 (2003).

\bibitem{Smith2003} D. R. Smith, D. Schurig, M. Rosenbluth, S. Schultz, S. A. Ramakrishna, and J. B. Pendry, Appl. Phys. Lett. \textbf{82}, 1506 (2003).

\bibitem{Kawata2001} S. Kawata (ed.), \textit{Near-Field Optics and Surface Plasmon Polaritons} (Springer, Berlin, 2001).

\bibitem{Chan1995} C. T. Chan, Q. L. Yu, and K. M. Ho, Phys. Rev. B \textbf{51}, 16635 (1995).

\bibitem{Taflove2000} A. Taflove and S. C. Hagness, \textit{Computational Electrodynamics: The Finite-Difference Time-Domain Method} (Artech House, Norwood, Mass., 2000).

\bibitem{Rao2003b} X. S. Rao and C. K. Ong, cond-mat/0304474.

\bibitem{Ruppin2000} R. Ruppin, Phys. Lett. A \textbf{277}, 61 (2000).

\bibitem{Ruppin2001} R. Ruppin, J. Phys.: Condens. Matter \textbf{13}, 1811 (2001).

\bibitem{Pendry2002} J. B. Pendry, cond-mat/0206561.

\end{thebibliography}
\end{document}